\documentclass[a4paper,12pt]{article}

\usepackage[utf8]{inputenc} 
\usepackage[francais]{babel}
\usepackage[OT1]{fontenc}       
\usepackage{graphicx}
\usepackage{times}

\usepackage{amsmath}
\usepackage{tikz}
\usetikzlibrary{positioning}
\usepackage{adjustbox}

\usepackage{fancyhdr}
\begin{document}

\renewcommand{\figurename}{Figure~}
\renewcommand{\tablename}{Tableau~}
\date{}

\title{ Vers une modélisation de la confiance dans le renseignement sur les menaces cyber \\[1.3em]
        Towards modeling trust in  Cyber Threat Intelligence 
}

\author{\small
     \begin{tabular}[t]{c@{\extracolsep{5em}}c@{\extracolsep{6em}}c}
        Laurent Bobelin${}^1$ & Sabine Frittella${}^1$  & Mariam Wehbe${}^1$ \\
        \end{tabular}
        ~\\
         ${}^1$        \small{INSA Centre Val de Loire}  
        ~\\
        \\
        \small{INSA Centre Val de Loire, 88 Bd Lahitolle, 18000 Bourges, France}\\
        \small{laurent.bobelin@insa-cvl.fr, sabine.frittella@insa-cvl.fr, mariam.wehbe@insa-cvl.fr}
}

\parskip 3mm
\maketitle

\thispagestyle{empty}

\begin{abstract}
    
Le renseignement sur les cybermenaces (\textit{Cyber Threat Intelligence}, ou CTI) est essentiel en cyberdéfense. Il recense les menaces qui peuvent peser sur un système informatique. Ces informations sont collectées par un agent par observation, ou en les recevant de sources. Il est nécessaire d'estimer la confiance dans chaque information acquise, en tenant compte des différentes dimensions qui peuvent composer la confiance : fiabilité de la source, compétence, plausibilité de l'information, crédibilité de l'information, par exemple. Ces informations recueillies doivent ensuite être agrégées pour consolider les renseignements. Des progrès ont été réalisés dans la théorie qui sous-tend la modélisation de la confiance quand elle est multidimensionnelle et que potentiellement certaines valeurs des dimensions sont indéterminées. Nous présentons ici la problématique du CTI et le partage de renseignements, et les raisons pour lesquelles nous utilisons une logique pour une première implémentation.
\end{abstract}

\section{Introduction}
\label{S:introduction}

Pour améliorer la sécurité des systèmes informatiques, une compréhension approfondie des risques spécifiques auxquels une organisation est confrontée est nécessaire. Avec une connaissance des menaces réelles, les organisations peuvent mieux se défendre. Percevoir, détecter et mitiger les vulnérabilités, prendre des mesures de sécurité adéquates pour améliorer leur posture de sécurité est le quotidien des équipes de cybersécurité.

Pour comprendre les cybermenaces et avoir une compréhension détaillée des risques, les organisations (entreprises, collectivités, ou toute autre entité devant défendre un système d'information) ont besoin de recueillir et exploiter les renseignements sur les cybermenaces (\textit{Cyber Threat Intelligence} ou CTI). Les renseignements collectés le sont généralement par des cellules spécialisées. Celles-ci vont de manière active récolter des informations, soit par :
\begin{itemize}
    \item la recherche dans les informations disponibles sur un sujet spécifique dans le domaine public (on parle alors d'\textit{Open Source INTelligence} ou OSINT),
    \item des méthodes actives : intrusion dans les systèmes des organisations attaquantes, espionnage ou infiltration,
    \item des observations (audit, analyse de logs, etc)
    \item la récolte de renseignements auprès de sources externes.
\end{itemize}

Les deux premières méthodes d'acquisition de renseignements sont l'apanage de grandes organisations (essentiellement les gouvernements par l'intermédiaire de leurs agences dédiées : CISA aux Etats-Unis, ANSSI, DGA, DGSI, DGSE en France) car elles demandent des ressources humaines spécialisées dans l'OSINT ou la recherche active d'information. De plus, parmi les méthodes de recherche, certaines impliquent de contrevenir à la loi ; de plus, le recel d'informations acquises illégalement est en France interdit par la loi, rendant complexe l'acquisition de données par le privé. 

Les deux dernières méthodes (observations et récoltes de renseignements externes) sont utilisées au quotidien par les équipes de cybersécurité pour les organisations privées. Les sources externes, dans le dernier cas, peuvent être des gouvernements et/ou des agences gouvernementales qui ont utilisé les deux premières méthodes pour récolter des informations, ou des acteurs privés. La confiance que l'on peut avoir dans ces informations est variable.


Ces informations sont utilisées par les organisations pour l'aide à la décision, notamment dans le déploiement de mesures de sécurité défensives ou pour stopper des activités qui peuvent paraître suspectes. Stopper des activités à tort peut causer des pertes financières, et paralyser les activités de l'entreprise. Il est donc essentiel de mettre au point des outils et des modèles permettant d'évaluer la confiance dans les informations reçues. Pour cela, il est nécessaire d'avoir un formalisme pour agréger des informations incertaines qui viennent de sources variées, et qui peuvent être possiblement paraconsistantes. 



L'objectif de cet article est de présenter la problématique de l'agrégation de renseignements cyber (CTI) et les caractéristiques clés qui nous ont poussé à choisir un modèle basé sur les logiques multivaluées pour mener des expériences pour une première implémentation.


\emph{Structure de l'article.} La section \ref{S:CTX} introduit le contexte du CTI, puis la section \ref{S:CTI} présente le CTI en lui-même : le processus de CTI est décrit dans la sous-section \ref{S:pCTI}, les informations qui sont produites par le CTI sont abordées dans la sous-section \ref{S:IoC},  la sous-section \ref{S:TLP} présente les standards et protocoles du partage d'information dans ce contexte, la sous-section \ref{S:STIX} le format de stockage de données.  La section \ref{S:SIEM} présente les outils intégrant les renseignements et les autres informations qui décrivent un système. Après avoir brièvement décrit notre problématique en section \ref{S:PB}, nous donnons dans la section \ref{S:implementation} les raisons qui nous ont poussé à choisir le modèle proposé dans \cite{Ch9,DBLP:conf/ipmu/dAllonnesL14, RevaultLesot2015} comme la base de première expérience dans ce domaine. 
Enfin, un état de l'art est donné dans la section \ref{S:RW}, et la section \ref{S:conclucion} conclut notre travail et examine les pistes de recherche futures.



\section{Cyberdéfense et décision}
\label{S:CTX}
De nos jours, toute organisation possédant un système informatique doit être capable de le défendre. Cela passe par plusieurs aspects distincts et complémentaires. Par exemple, on peut par construction rendre un système plus sûr, en utilisant des méthodes de conception, d'implémentation et de déploiement particulières.

Néanmoins, on ne peut se prémunir de toutes les attaques par construction. Il convient, lorsque le système est opérationnel, de défendre le système contre les menaces : pour cela, on forme une équipe d'experts chargés de défendre le système, appelée \textit{blue team} qui travaille dans un \textit{Security Operation Center (SOC)}. Celle-ci doit être capable de détecter et d'anticiper d'éventuelles attaques, de décider de contre mesures en cas d'attaque, de les appliquer et d'évaluer leur efficacité.  

\begin{figure} [htbp]
    \centering
    \includegraphics[width=0.34\paperwidth]{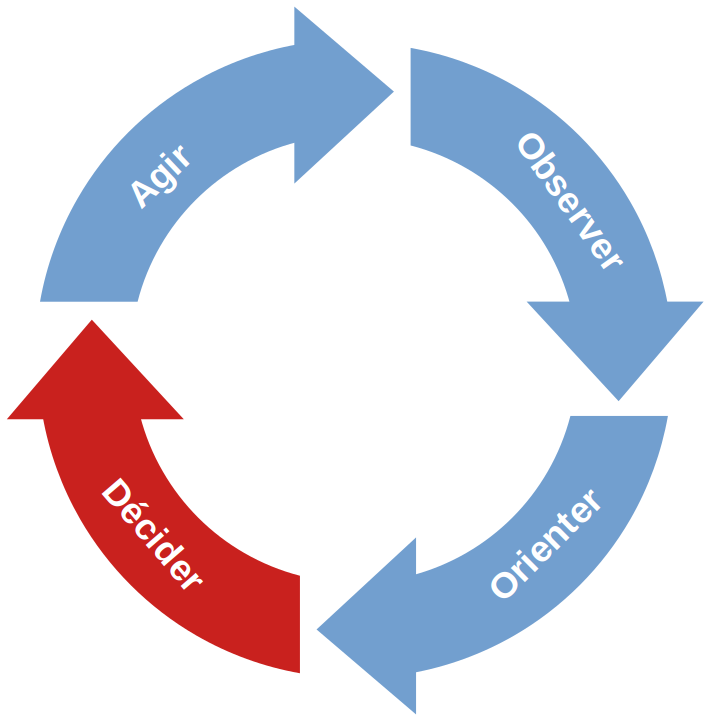}
    \caption{La boucle OODA}
    \label{fig:OODA}
\end{figure}

Une modélisation populaire de ce processus de prise de décision est la boucle OODA (Observer, Orienter, Décider, Agir) \cite{Clarke2019-my} présentée dans la figure \ref{fig:OODA}. Ce modèle permet de réagir rapidement en collectant et en analysant les informations dans des environnements de cybersécurité où le temps de réponse aux menaces est limité. La phase \textbf{Observer} consiste pour une organisation à surveiller son environnement et celui de ses adversaires pour détecter des signes de menace potentielle. La phase \textbf{Orienter} permet de contextualiser et analyser les informations recueillies dans la phase précédente pour identifier les tendances d'attaque. Potentiellement, cette phase peut demander la recherche d'informations plus précises sur une partie de l'environnement (capacité des attaquants, compromission de machine par exemple). À la lumière de ces analyses, des décisions sont prises dans la phase \textbf{Décider} pour répondre aux menaces identifiées. Enfin, ces décisions se concrétisent en actes pour contrer les menaces dans la phase \textbf{Agir}.

\section{Renseignement sur les cybermenaces (CTI)}
\label{S:CTI}
La boucle OODA repose donc sur des informations sur l'environnement, qui sont la résultante de la phase d'observation. Elles sont généralement récoltées de manière passive (réception d'information diffusée par des tiers). Cette boucle peut nécessiter la production d'information par l'organisation lors de la phase d'orientation. Ces informations sont produites par le processus du CTI \cite{221741}. 

Pour produire les informations, une organisation (par l'intermédiaire de ses agents) utilise des sources internes et externes pour recueillir diverses informations sur la menace et de tirer parti de son analyse. Cela permet de construire une connaissance qui sera communiquée entre autres aux équipes de cybersécurité (blue team). Ces informations seront exploitées pour la défense du système d'information de l'organisation, que ce soit pour la détection d'intrusion, ou le déploiement de contre-mesures (arrêt/interdiction de flux de données, bannissements de plages d'adresses par exemple). 

\subsection{Le processus du CTI}
\label{S:pCTI}
\begin{figure} [htbp]
    \centering
    \includegraphics[width=0.4\paperwidth]{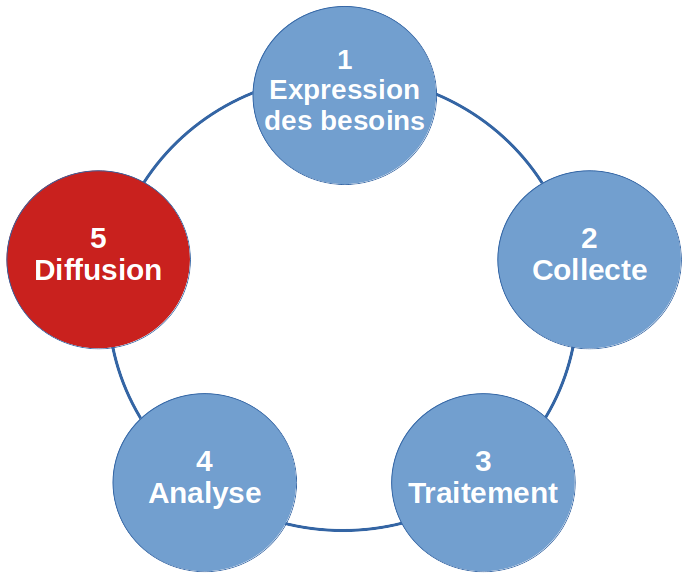}
    \caption{Cycle du renseignement CTI}
    \label{fig:CTI}
\end{figure}

Pour gérer les opérations d'une activité de renseignement, les organisations appliquent \textit{un processus CTI}, qui leur permet de structurer leur acquisition de renseignement. Une modélisation répandue de ce processus est un cycle comprenant cinq phases \cite{zvelo, Montasari2021} : planification et orientation, collecte, traitement, analyse et diffusion comme le montre la figure \ref{fig:CTI}. La phase de \textbf{planification et orientation} permet à partir d'une expression des besoins de spécifier quel type d'informations les organisations souhaitent obtenir et quels sont les moyens utilisés pour les obtenir. Par exemple, un état peut décider de surveiller plus spécifiquement des groupes de hackers originaires d'un pays particulier à l'approche d'un évènement potentiellement à risque (Jeux Olympiques par exemple). Ensuite, la phase de \textbf{Collecte} permet de collecter des données techniques pour détecter les attaques en cours et tenter d'anticiper les prochaines. Ces méthodes peuvent être actives - par l'infiltration de groupes, l'OSINT ou l'accès aux ressources de l'adversaire, ou semi-passives par l'utilisation de \textit{honeypots}\footnote{Les \textit{honeypot} sont des ressources (site Web, VM, réseau) délibérément rendues vulnérables à certaines attaques et instrumentées pour pouvoir récolter de l'information sur les potentiels attaquants.} ou la demande d'information auprès de sources. Ces données brutes seront converties en informations utiles durant la phase de \textbf{Traitement} afin de connaître les risques de sécurité et détecter les attaques. La phase \textbf{Analyse} permet de synthétiser les informations en renseignements qui aident à prendre une décision et à mener des actions préventives. Elle comprend l'analyse des modèles, l'évaluation et la notation qui répondent aux besoins spécifiés dans la phase de planification. Finalement, la phase de \textbf{diffusion}  consiste à partager ces informations avec des tiers ; la diffusion se fait suivant des politiques de partage. Elles prennent en compte le degré de criticité des informations et la nature des relations qui lient l'organisme diffuseur et ceux qui reçoivent ces informations.

La liaison entre le cycle de prise de décision et celui de CTI se fait à deux niveaux : 
\begin{itemize} 
\item L'expression des besoins à la base de la phase de planification  et d'orientation du processus CTI est la résultante de la phase d'orientation de la boucle OODA.
\item les informations diffusées lors de la phase idoine du CTI sont celles qui sont observées lors de la phase Observer de la boucle OODA. 
\end{itemize}

Le produit du processus CTI est un ensemble de renseignements ; un des types de renseignements les plus classiques produits est l'indicateur de compromission (\textit{Indicator of Compromise} ou IoC).

\subsection{Les informations du renseignement (IoC)}
\label{S:IoC}
La phase d'analyse du processus CTI peut produire un indicateur de compromission. Une compromission se produit lorsqu'un attaquant prend le contrôle d'une machine en y introduisant un logiciel malveillant (malware). Une fois le malware en place, l'attaquant peut exécuter des actions malveillantes telles que voler des données sensibles, espionner les activités de l'utilisateur, ou utiliser la machine compromise pour lancer des attaques supplémentaires sur d'autres systèmes. L'adresse IP source d'une attaque et la signature d'un fichier exécutable malveillant sont des exemples typiques d'IoCs \cite{GUARASCIO202230}. 



Il est important de noter qu'un IoC n'est pas l'affirmation certaine que l'on est en face d'une compromission, mais la combinaison d'un ensemble de preuves, permettant de penser avec un certain degré de certitude qu'une observation similaire à l'IoC indique que l'on est en face d'une compromission. Plus précisément, cela signifie que la présomption de compromission est quelque chose d'incertain.  

\subsection{Politique de diffusion des renseignements}
\label{S:TLP}

Lors de la phase de diffusion du processus CTI, l'équipe de renseignement diffuse l'information suivant une politique qui lui permet de choisir quelles informations seront diffusées, et à qui. En effet, les informations sont potentiellement sensibles, et leur divulgation trop large pourrait entrainer une perte de valeur de l'information elle-même. Par exemple, révéler à un attaquant que l'on sait quelles sont ses méthodes d'attaques pourrait le pousser à changer de méthode, et donc rendre l'information sans intérêt. Il convient donc de restreindre la diffusion de cette information.

\begin{figure} [htbp]
    \centering
    \includegraphics[width=0.4\paperwidth]{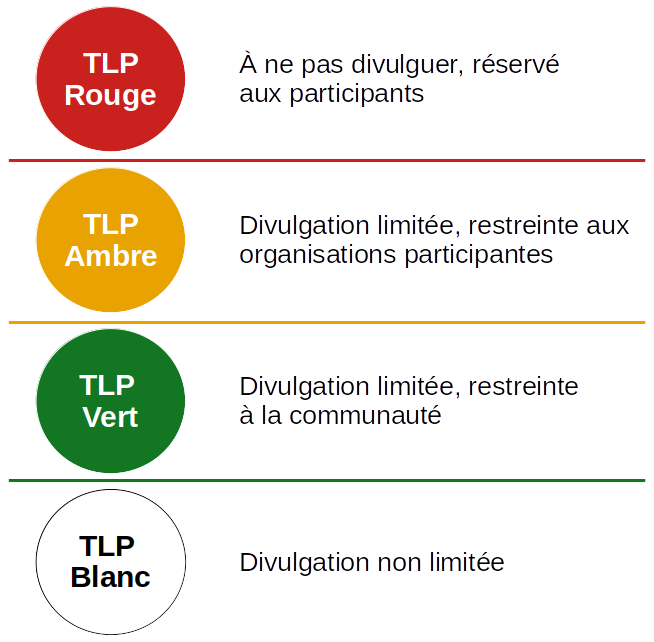}
    \caption{Traffic Light Protocol TLP}
    \label{fig:TLP}
\end{figure}

Le partage d'IoCs respecte donc un ensemble de lignes directrices visant à garantir que les informations sensibles sont partagées avec le public approprié. Le protocole Traffic Light Protocol (TLP) est communément utilisé pour définir les modes de diffusion. TLP a été créé pour déterminer la confidentialité des informations suivant les niveaux décrits à la figure \ref{fig:TLP}. TLP 
 considère que l'on transmet des informations à des \textit{participants} qui appartiennent à des \textit{organisations}, qui elles-mêmes font partie de \textit{communautés}, et définit ces niveaux en fonction de la diffusion des informations. \textbf{Rouge} est le plus haut niveau de confidentialité des informations, qui doivent être partagées seulement avec des participants précis. L'\textbf{Ambre} permet la divulgation à l’intérieur de l’organisation des participants. Alors que le \textbf{Vert} gère les informations réservées à la communauté. Enfin, le \textbf{Blanc} permet une diffusion publique sans restriction.

\subsection{Représentation des connaissances sur les menaces}
\label{S:STIX}
\begin{figure} [htbp]
    \centering
    \includegraphics[width=0.38\paperwidth]{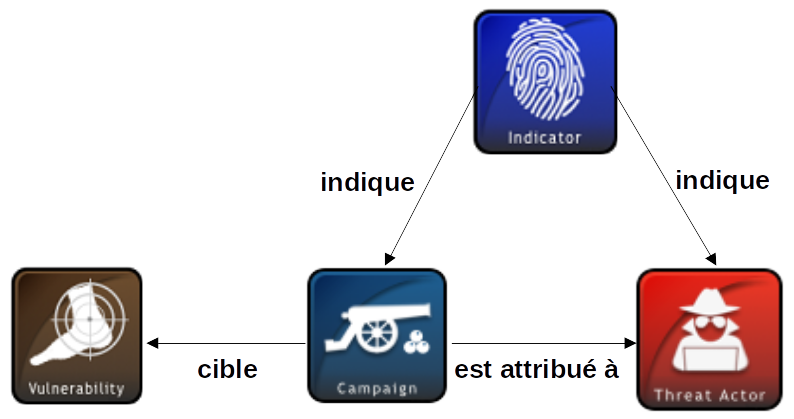}
    \caption{STIX}
    \label{fig:STIX}
\end{figure}

Ces informations diffusées sont ensuite intégrées dans des plateformes CTI, qui reposent souvent sur des ontologies. Le langage Structured Threat Information Expression (STIX), reposant sur une ontologie, est le standard de facto dans le CTI. Cela facilite leur lecture par l'humain comme la machine, ainsi que l'analyse des informations et le partage de CTI entre les organisations, ce qui permet d'améliorer la capacité des équipes de cybersécurité à prendre des mesures de sécurité, à détecter et mitiger les menaces connues rapidement. On donne comme exemple la modélisation d'une activité malveillante décrite dans la figure \ref{fig:STIX}. Les différents objets présents dans ce scénario (indicateur, acteur de menace, campagne et vulnérabilité), et qui sont offerts par l'ontologie STIX, indiquent l'identité de l'attaquant, la campagne d'attaque utilisée et les vulnérabilités identifiées pour l'attaque. 

\section{Observations de l'état du système (SIEM)}
\label{S:SIEM}
Le CTI n'est pas la seule source d'information à la base de la prise de décision de la boucle OODA. La \textit{blue team} possède un ensemble d'informations sur son environnement, souvent des informations dites de \textit{monitoring}, qui combine les informations de divers logs d'activité, de sondes potentiellement déployées sur des machines et des réseaux pour détecter des signatures particulières d'activité. Les informations sont regroupées au sein d'un outil appelé systèmes de gestion des informations et des événements de sécurité ou Security Information and Event Management (SIEM).

Les SIEM jouent un rôle crucial dans la protection des organisations contre les cyberattaques sophistiquées. Ces solutions centralisent et analysent les données de sécurité provenant de diverses sources, telles que les pare-feu, les serveurs, les points d'extrémité et les réseaux, offrant une visibilité globale sur l'activité de sécurité au sein de l'infrastructure informatique.

En corrélant les événements de sécurité disparates, possiblement les informations de CTI, les SIEM permettent aux équipes de sécurité d'identifier rapidement les anomalies et les comportements suspects qui pourraient indiquer une activité malveillante. Cette capacité de détection précoce est essentielle pour contenir les cyberattaques et minimiser les dommages potentiels.

Les SIEM offrent également des fonctionnalités avancées d'investigation et de réponse aux incidents, permettant aux analystes de sécurité de retracer l'origine d'une attaque, d'identifier les systèmes compromis et de prendre les mesures correctives appropriées. De plus, les SIEM peuvent être intégrés à d'autres solutions de sécurité, telles que les systèmes de prévention des intrusions ou Intrusion Prevention System (IPS) et les systèmes de détection des intrusions ou Intrusion Detection System (IDS), pour automatiser la réponse aux menaces.

L'adoption de SIEM est devenue une pratique standard pour les organisations de toutes tailles qui cherchent à renforcer leur posture de sécurité et à se protéger contre les cybermenaces en constante évolution. En offrant une visibilité et une analyse centralisées des données de sécurité, les SIEM permettent aux équipes de sécurité de détecter, d'enquêter et de répondre aux incidents de manière plus efficace, réduisant ainsi le risque de compromission et les impacts négatifs sur les opérations commerciales. 

\section{Problématique}
\label{S:PB}
Dans le cadre de l'OODA, nous nous intéressons plus particulièrement à la modélisation de la confiance dans les informations qui sont utilisées lors de la phase d'observation. Dans un premier temps, nous avons choisi de nous restreindre aux informations de CTI.

Ces informations sont structurées sous forme de flux en provenance de sources, et formatées sous la forme d'ontologie STIX, le plus souvent. Des outils comme OpenCTI \cite{opencti} recommandés par l'Agence Nationale de la Sécurité des Systèmes d'Information ou National Agency for the Security of Information Systems (ANSSI) permettent d'agréger les différents flux sous la forme d'une ontologie unique. 

Les informations que l'on récolte proviennent en pratique de différentes sources, qui vont choisir de nous diffuser des informations avec un certain niveau de confidentialité. L'organisation qui utilise ces informations accorde des degrés de fiabilité divers dans les différentes sources, qui dépendent des relations entre l'organisation et les sources. 

La problématique que nous voulons adresser est la manière d'agréger des informations  en prenant en compte différents aspects liés à la confiance dans ces informations, ainsi qu'à l'incertitude intrinsèque de l'information elle-même. Les informations peuvent être paraconsistantes. Certaines dimensions de la confiance (fiabilité, compétence, plausibilité, crédibilité) peuvent ne pas être connues. 

\section{Implémentation}
\label{S:implementation}

Évaluer la confiance dans l'information nécessite un formalisme pour agréger des informations incertaines provenant de diverses sources. Étant donné que le modèle proposé dans \cite{Ch9,DBLP:conf/ipmu/dAllonnesL14, RevaultLesot2015} offre une modélisation de la confiance pour l'aide à la décision en se basant sur des informations incertaines et incomplètes, nous l'avons implémenté comme base pour une première expérience dans ce domaine afin de réaliser des tests. Ce modèle présente un cadre logique multivaluée basé sur les dimensions suivantes : la fiabilité et la compétence de la source, la plausibilité et la crédibilité de l'information reçue. Ce modèle est utilisé pour encoder les informations disponibles sur chaque dimension. Par exemple, si une source annonce "l'hôte A est compromis", pour évaluer la crédibilité de cette information, nous évaluerons tout d'abord le degré de fiabilité et de compétence de la source, puis sa plausibilité par rapport à la connaissance de l'agent et ensuite sa crédibilité par rapport à d'autres informations. La confiance est calculée sur la base d'une agrégation des différentes dimensions de qualité de la source et de l’information. Dans ce contexte, nous avons mis en œuvre ce modèle théorique en langage de programmation Java et avons développé un outil de construction de la confiance en se basant sur la logique multivaluée ainsi que l'implémentation des opérateurs proposés et les matrices définissant les stratégies d'évaluation de la confiance. L'intégration de la confiance dans les ontologies étant un sujet complexe, nous avons choisi comme premiers travaux de ne pas considérer les ontologies dans leur ensemble, mais uniquement des IoC les plus simples possibles (du type "L'hôte A est compromis").

\section{État de l'art}
\label{S:RW}
L'étude du CTI est actuellement en plein essor dans la communauté de la cybersécurité, mais pour l'instant peu de travaux portent sur la modélisation formelle de la problématique posée par la confiance dans les sources d'information. Beaucoup d'articles sont en effet tournés vers les aspects pratiques, l'étude du domaine étant principalement porté par la pratique et l'industrie. Pour avoir une vue d'ensemble de l'état des connaissances dans le domaine, on peut par exemple se référer à \cite{Saeed2023} qui est une revue de la littérature systématique sur le CTI pour l'aide à la décision, ou bien \cite{ALAEIFAR2024103786} qui est un état de l'art sur le partage d'information de CTI. Ce dernier identifie la gestion de l'incertitude (épistémique et/ou aléatoire) comme un challenge futur dans le domaine. Une étude des différents protocoles et formats d'échange \cite{electronics9050824} met aussi en lumière que si le standard de facto est STIX, il n'existe aucune alternative prenant en compte de tels aspects. 

\section{Conclusion}
\label{S:conclucion}
Dans cet article, nous avons décrit le processus de renseignement sur les menaces et son intégration dans un processus décisionnel. Ces informations CTI sont par nature imprécises, de sources plus ou moins fiables, et comptent parmi les informations nécessaires à la prise de décision dans des environnements de cybersécurité. 

En partant de l'analyse de la nature des informations que l'on partage, nous avons choisi pour une première expérimentation  un modèle théorique basé sur une logique multivaluée. 
Notre but est d'explorer et d'estimer, par des études de cas et des expériences, la pertinence du modèle dans notre contexte. A terme, nous aimerions introduire cette confiance lors de l'agrégation de modèles de données plus complexes, comme par exemple les ontologies de type STIX, et donc d'introduire cette notion dans les outils de CTI.

\section*{Remerciements}
Les travaux de Sabine Frittella sont financés par l'ANR JCJC 2019, projet PRELAP (ANR-19-CE48-0006). 
Les travaux de Mariam Wehbe sont financés par Projet CyberINSA France 2030 ANR-23-CMAS-0019\footnote{https://cyberinsa.insa-cvl.fr/}.
Ces travaux rentrent dans le cadre du projet MOSAIC financé par l'Union Européenne, bourse Marie Sk\l{}odowska-Curie No.~101007627.


\end{document}